\documentclass{PoS}

\usepackage{bm}

\newcommand\onelineequation[2]{%
\begin{equation}\label{#1}
#2
\end{equation}%
}
\newcommand\onefigure[4]{%
\begin{figure}[#4]
\centering
\includegraphics[width=#2]{#1}
\caption{#3}\label{#1}
\end{figure}%
}
\newcommand\twofigures[5]{%
\begin{figure}[#5]
\centering
\begin{tabular}{c c}
\includegraphics[width=#3]{#1}&\includegraphics[width=#3]{#2}
\end{tabular}
\caption{#4}\label{#1_#2}
\end{figure}%
}
\newcommand\oh{{\textstyle\frac{1}{2}}} 
\newcommand\of{{\textstyle\frac{1}{4}}} 
\newcommand\cN{{\cal{N}}}
\title{Measuring the ground-state wave functional\\of SU(2) Yang--Mills theory in 3\,+\,1 dimensions\\{\Large Abelian plane waves}%
\addtocounter{footnote}{1}\thanks{This research was supported in part by the U.S.\ DOE under Grant No.\ DE-FG03-92ER40711 (J.G.), by the Slovak Research and Development Agency under Contract No.\ APVV--0050--11, and by the Slovak Grant Agency for Science, Project VEGA No.\ 2/0072/13 (\v{S}.O.). In initial stages of this work,  \v{S}.O.\ was also supported by ERDF OP R\&D, Project meta-QUTE ITMS 2624012002.}}

\ShortTitle{Measuring the ground-state wave functional of SU(2) YM theory\dots} 

\author{Jeff Greensite\\
Physics and Astronomy Dept., San Francisco State University, San Francisco, CA 94132, USA\\
E-mail: \email{jgreensite@gmail.com}}

\author{\addtocounter{footnote}{-2}
				\speaker{\v{S}tefan Olejn{\'\i}k}\\
        Institute of Physics, Slovak Academy of Sciences, SK--845 11 Bratislava, Slovakia\\
        E-mail: \email{stefan.olejnik@savba.sk}} 

\abstract{A method of measuring relative probabilities of various gauge-field configurations in the Yang--Mills vacuum was proposed long ago [Phys.\ Lett.\ B \textbf{223} (1989) 207]. We applied this method to compute the square of the YM vacuum wave functional (VWF) in numerical simulations of SU(2) lattice gauge theory in $D=3+1$ dimensions for sets of abelian plane waves. The results were compared to predictions based on some VWF proposals in the literature. None of them describes the data satisfactorily at large plane-wave momenta. The phenomenological best fit to data, containing three free parameters, appears to reduce in the continuum limit to the approximate form proposed in [Phys.\ Rev.\ D {\bf 77} (2008) 065003].}

\FullConference{31st International Symposium on Lattice Field Theory LATTICE 2013\\
		 July 29 -- August 3, 2013\\
		 Mainz, Germany}

\begin{document}

\section{Formulation of the problem}\label{section1}
One looks for the vacuum wave functional satisfying the Schr\"odinger equation of the SU(2) gauge theory in temporal gauge:
\onelineequation{SchR}{\int d^3x\left(-\oh\frac{\delta^2}{\delta A^a_k(x)^2}+\of F^a_{ij}(x)^2\right)\Psi_0[A]=E_0\Psi_0[A].}%
Solutions of the equation corresponding to physical states must satisfy the Gau\ss-law constraint:
\onelineequation{Gauss}{\left(\delta^{ac}\partial_k+g\varepsilon^{abc} A^b_k(x)\right)\frac{\delta\Psi[A]}{\delta A^c_k(x)}=0.}%
The problem of finding the Yang--Mills VWF has been addressed by various techniques, in temporal and Coulomb gauges, in 2 and 3 space dimensions, but is still far from a full solution. (For a~brief review and references see \textit{e.g.\/} Sec.\ II of Ref.~\cite{Greensite:2011pj}, the most recent work has been published in Ref.~\cite{Krug:2013yq}.) Relative weights of various gauge-field configurations can be computed also numerically, at least for some simple subsets, in simulations of the Yang--Mills theory in the lattice formulation, and we will present here results for sets of abelian plane waves and compare them to expectations based on a few proposed forms of the VWF. 

\section{Proposals of the SU(2) Yang--Mills vacuum wave functional}\label{section2}
	On the basis of strong coupling and the idea of a magnetically disordered vacuum, one of us (J.G.)~\cite{Greensite:1979yn} suggested long ago that the VWF in $(3+1)$ dimensions might assume the following form: 
\onelineequation{DR}{\Psi_0[A]=\cN\exp\left(-\oh\mu\int d^3x\;\mbox{Tr}[F^2_{ij}(x)]\right)\quad\dots\quad\underline{\mbox{DR}}}%
at least for sufficiently long-wavelength, slowly varying configurations.  A similar suggestion in $(2+1)$ dimensions was made by Halpern~\cite{Halpern:1978ik}. This so-called \textit{dimensional-reduction} form cannot of course be right for \textit{all} gauge fields, since it has some unwanted consequences, \textit{e.g.\/} exact Casimir scaling and incorrect short-distance behaviour.

	We proposed another approximate VWF~\cite{Greensite:2007ij}, that interpolates between the known $g\to0$ limit and the dimensional-reduction form,
\onelineequation{GO}{\Psi_0[A]=\cN\exp\left[-\of\int d^3x\;d^3y\;F^a_{ij}(x)\left(\frac{1}{{\cal{K}}[A;m]}\right)^{ab}_{xy} F^b_{ij}(y) \right]\quad\dots\quad\underline{\mbox{GO}}}%
where
\onelineequation{kernel}{{\cal{K}}[A;m]\equiv-{\cal{D}}^2[A]-\lambda_0+m^2,}%
${\cal{D}}^2[A]$ is the covariant laplacian in the adjoint representation, shifted by $\lambda_0$, the lowest eigenvalue of $(-{\cal{D}}^2[A])$, and $m$ is a free (mass) parameter. This form, a variation on an earlier VWF proposal of Samuel~\cite{Samuel:1996bt}, was shown to be a fairly good approximation to the true ground state of the theory in $(2+1)$ dimensions, see \cite{Greensite:2011pj,Greensite:2007ij,Greensite:2010tm} for details.

	An ambitious attempt to compute the VWF analytically in $(2+1)$ dimensions was undertaken by Karabali, Kim, and Nair~\cite{Karabali:1998yq}. They rewrote the theory with help of new gauge-invariant variables, and derived an expression for the VWF in 
their terms. When expressed again in the old variables, their VWF assumes the form:
\onelineequation{KKNngi}{\Psi_0[A]=\cN\exp\left[-\oh\int d^2x\;d^2y\;B^a(x)\left(\frac{1}{\sqrt{-\nabla^2+m^2}+m}\right)_{xy} B^b(y) \right],}
which, however, is \textit{not} gauge-invariant. A gauge-invariant generalization to $(3+1)$ dimensions, along the lines of the proposal (\ref{GO}), might be
\onelineequation{KKN3}{\Psi_0[A]=\cN\exp\left[-\of\int d^3x\;d^3y\;F^a_{ij}(x)\left(\frac{1}{{\cal{K}}[A;m]+m}\right)^{ab}_{xy} F^b_{ij}(y) \right]\quad\dots\quad\underline{\mbox{KKN}}}
This form will also be below confronted, for comparison's sake, with numerical data.  However, we stress that this is not a proposal by the authors of Ref.~\cite{Karabali:1998yq},  but is only a conjecture inspired by (\ref{GO})--(\ref{KKNngi}). 

\section{Measurement method}\label{section3}
The squared VWF is given (on a lattice in the temporal gauge) by the path integral:
\onelineequation{PI}{\Psi^2_0[U']=\frac{1}{Z}\int [DU]\delta(U_0)\prod_{\mathbf{x},i}\delta[U_i(\mathbf{x},0)-U'(\mathbf{x})]e^{-S[U]}.}%
\noindent
The \textit{{relative-weight method}} \cite{Greensite:1989aa} enables one to compute ratios $\Psi^2[U^{(n)}]/\Psi^2[U^{(m)}]$  for configurations belonging to a finite set
${\mathcal{U}}=\lbrace U_i^{(j)}(\mathbf{x}),j=1,2,\dots,M\rbrace$ (assuming they are nearby in the configuration space). One uses Monte Carlo simulations with the usual update algorithm (\textit{e.g.\/}\ heat-bath) for all spacelike links at $t\ne0$ and for timelike links, while the spacelike links at $t=0$ are updated all at once selecting one configuration from the set $\mathcal{U}$ at random and accepting/rejecting it via the Metropolis algorithm. Then\
\onelineequation{ratio}{\frac{\Psi^2[U^{(n)}]}{\Psi^2[U^{(m)}]}=\lim_{N_\mathrm{tot}\to\infty}\frac{N_n}{N_m}
=\lim_{N_\mathrm{tot}\to\infty}\frac{N_n/N_\mathrm{tot}}{N_m/N_\mathrm{tot}},}%
where $N_n$ ($N_m$) is the number of times the $n$-th ($m$-th) configuration is accepted and $N_\mathrm{tot}$ is the total number of updates.

	The VWF can be written in the form 
\onelineequation{}{\Psi^2[U]={\mathcal{N}}e^{-R[U]},}%
hence the measured values of $-\log(N_n/N_\mathrm{tot})$ should fall on a straight line with unit slope as function of $R[U^{(n)}]$, see Fig.~\ref{prob_k_2_3_l20} for illustration.
\onefigure{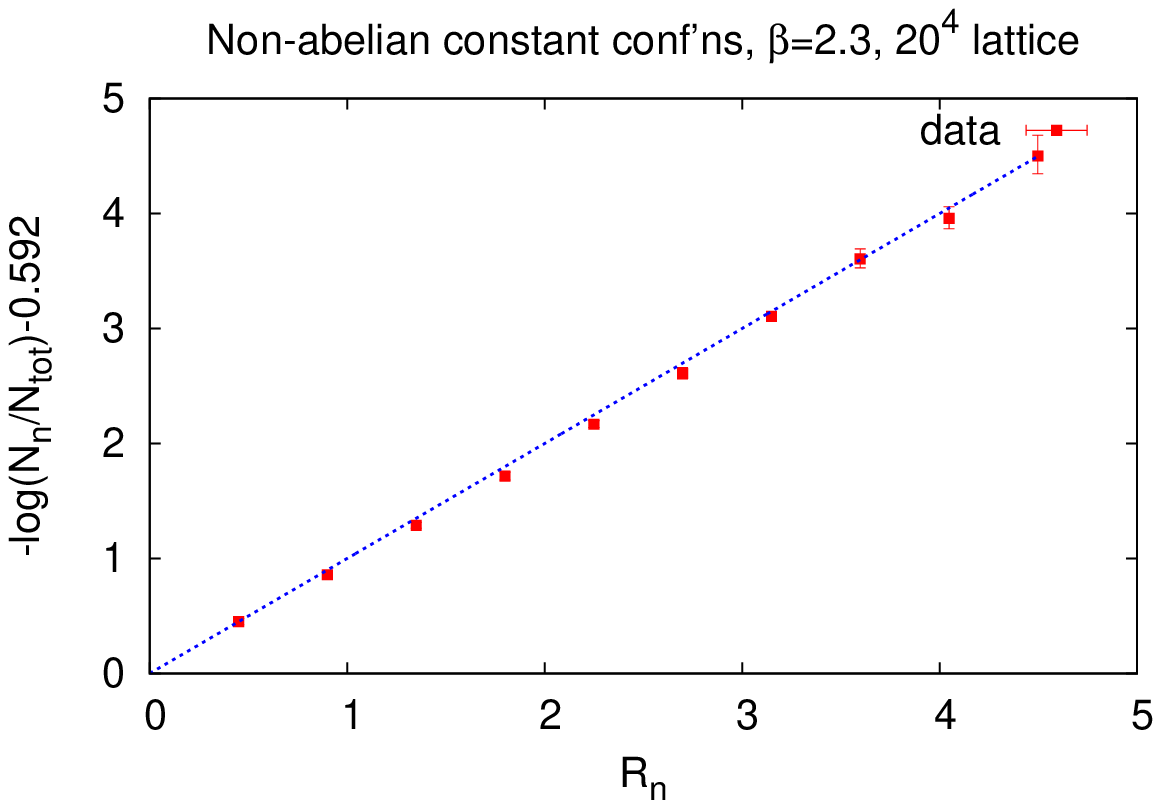}{0.5\textwidth}{$-\log(N_n/N_\mathrm{tot})$ (shifted by a constant) vs.\ $R_n=\mu\kappa n$ for ${\mathcal{U}}_\mathrm{nac}$ [see Eq.~(\protect\ref{nac})] with $\kappa=0.14$. The value of $\mu$ comes out $3.21 (5)$.}{tbh}

\section{Sets of test configurations}\label{section4}
	Until now, we have performed numerical simulations using the relative-weight method for two kinds of simple gauge-field configurations:
\begin{itemize}
\item
\textit{Non-abelian constant configurations}:
\onelineequation{nac}{{\cal U}_\mathrm{nac}=\left\{U_k^{(n)}(x)=\sqrt{1-\left(a^{(n)}\right)^2}\mathbf{1}+ia^{(n)}\bm{\sigma}_k\right\},} 
where
\onelineequation{nac_a}{a^{(n)}=\left(\frac{\kappa}{6L^3}n\right)^{1/4},\qquad n=1,2,\dots, 10.}%
	These configurations were used only to ``calibrate'' our computer code by comparison with the results of Ref.~\cite{Greensite:1989aa}, obtained on lattices of much smaller size, see \cite{Greensite:2013nb}.
\item
\textit{Abelian plane-wave configurations}: 
\onelineequation{apw}{{\cal U}_\mathrm{apw}=\left\{U_1^{(j)}(x)=\sqrt{1-\left(a^{(j)}_\textbf{\textit{n}}(x)\right)^2}\mathbf{1}+ia^{(j)}_\textbf{\textit{n}}(x)\bm{\sigma}_3,\quad U_2^{(j)}(x)=U_3^{(j)}(x)=\mathbf{1}\right\}, }
where $\textbf{\textit{n}}=(n_1,n_2,n_3)$, and
\onelineequation{apw_a}{a^{(j)}_\textbf{\textit{n}}=\sqrt{\frac{\alpha_\textbf{\textit{n}}+\gamma_\textbf{\textit{n}}j}{L^3}}\cos\left(\frac{2\pi}{L}\textbf{\textit{n}}\cdot\textbf{\textit{x}}\right),\qquad j=1,2,\dots, 10.}
	Pairs of $(\alpha_\textbf{\textit{n}},\gamma_\textbf{\textit{n}})$ characterizing abelian plane waves with the wavenumber $\textbf{\textit{n}}$ in the above equations were carefully selected so that the actions of plane waves with different $j$ were not much different (to ensure reasonable Metropolis acceptance rates in the method of Sec.~\ref{section3}).
\end{itemize}


\section{Showcase of results}\label{section6}
	For a particular set of abelian plane waves with the wavenumber $\textbf{\textit{n}}$  the measured values of relative weights of individual plane waves are expected to be a linear function of $(\alpha_\textbf{\textit{n}}+\gamma_\textbf{\textit{n}}j)$. By a fit of the form
\onelineequation{linear}{-\log(N^{(j)}_\textbf{\textit{n}}/N_\mathrm{tot})=R^{(j)}_\textbf{\textit{n}}=\oh(\alpha_\textbf{\textit{n}}+\gamma_\textbf{\textit{n}}j)\times{\omega(\textbf{\textit{n}})}}%
one can determine the slope $\omega(\textbf{\textit{n}})$. The inferred dependence of $\omega(\textbf{\textit{n}})$  on $\textbf{\textit{n}}$ can then be compared with expectations based on the DR, GO, and KKN-inspired vacuum wave functionals. In particular,
we performed the following fits:
\onelineequation{fits}{{\omega(\textbf{\textit{n}})}=\left\{
\begin{array}{l c l}
{b}k^2(\textit{\textbf{n}}) & \qquad\dots\qquad & \underline{\mbox{DR}},\\[2mm]
{\displaystyle{c}\frac{k^2(\textit{\textbf{n}})}{\sqrt{k^2(\textit{\textbf{n}})+{m}^2}}} & \dots & \underline{\mbox{GO}},\\[2mm]
{\displaystyle{c}\frac{k^2(\textit{\textbf{n}})}{\sqrt{k^2(\textit{\textbf{n}})+{m_1}^2}+{m_2}}} & \dots & \underline{\mbox{inspired by KKN}},
\end{array}\right.}%
where
\onelineequation{momentum}{k^2(\textit{\textbf{n}})=2\sum_i\left(1-\cos\frac{2\pi n_i}{L}\right).}

\twofigures{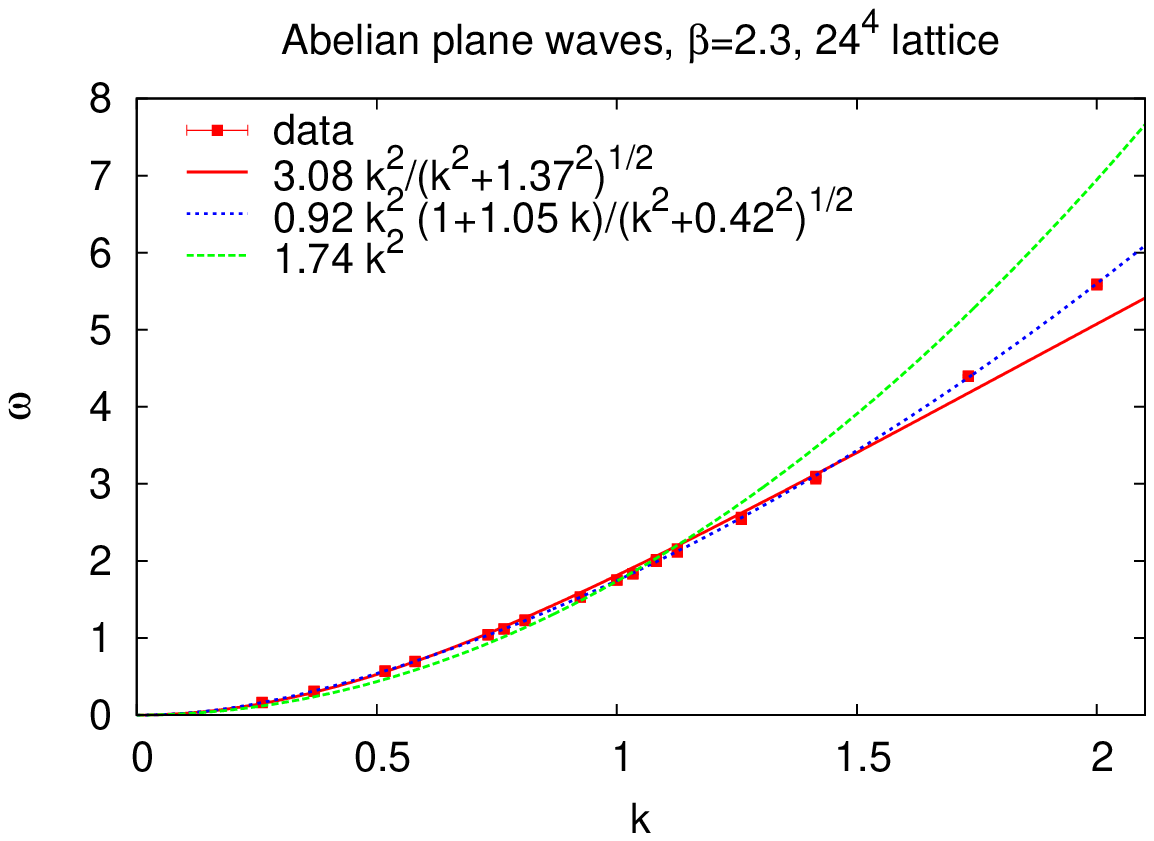}{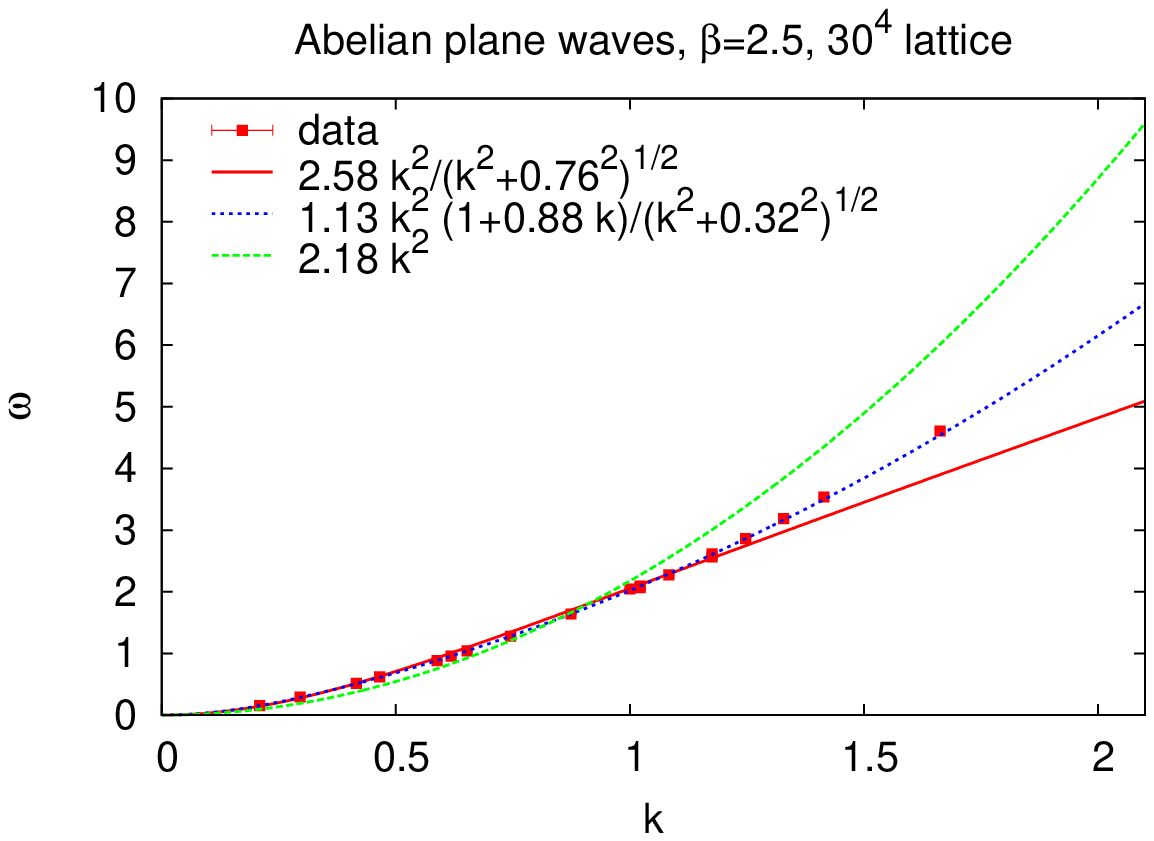}{0.48\textwidth}{Sample plots of $\omega(\textbf{\textit{n}})$ vs.\ $k(\textbf{\textit{n}})$ for ${\cal U}_\mathrm{apw}$ sets, with fits of the DR (green line) and GO (red line).}{t!}
\onefigure{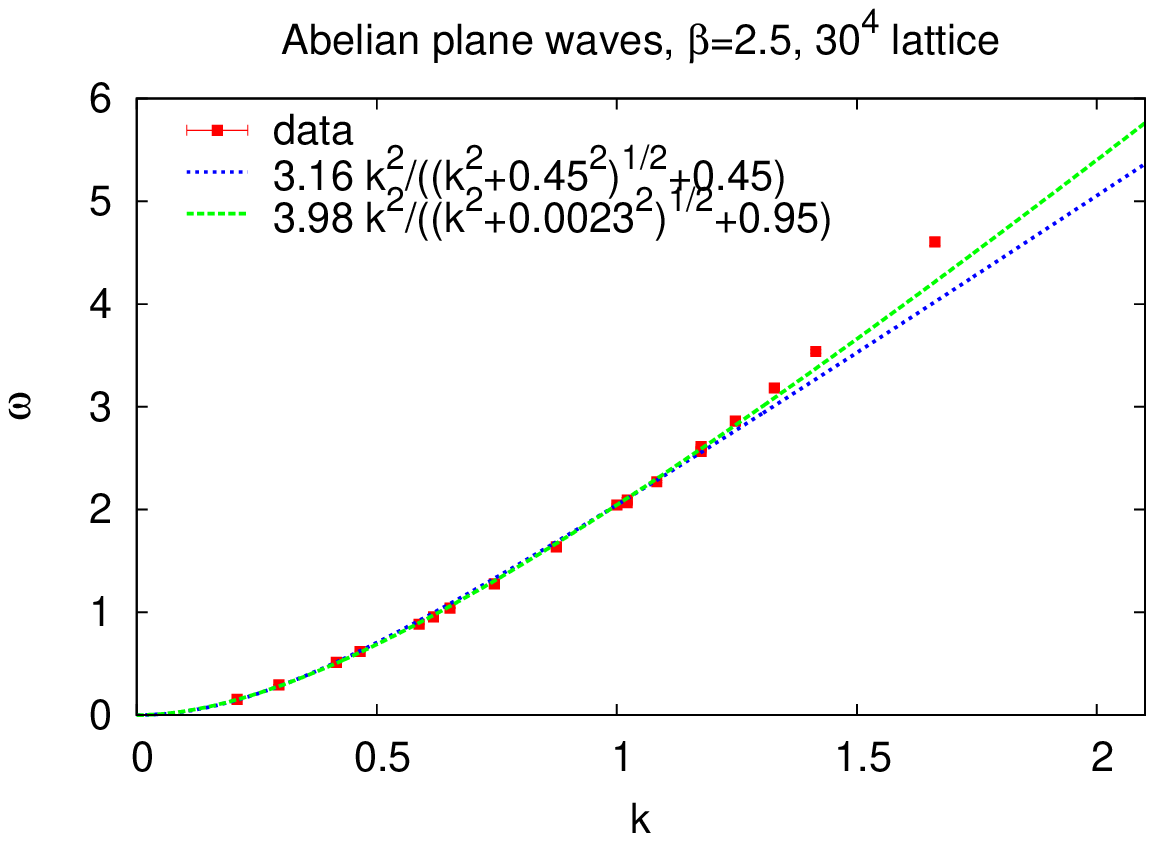}{0.5\textwidth}{ A sample plot of $\omega(\textbf{\textit{n}})$ vs.\ $k(\textbf{\textit{n}})$ for ${\cal U}_\mathrm{apw}$, with ``KKN-inspired'' fits.}{b!}

	In the KKN-inspired fit we used two mass parameters, $m_1$ and $m_2$, instead of just $m$, \textit{cf.}~Eq.~(\ref{KKN3}). We then performed one fit with both parameters free, and another constrained fit with $m_1=m_2$. It turned out that the former had a lower $\chi^2$ and the preferred value of $m_1$ was close to 0.

	The sample plots for fits of the form (\ref{fits}) are displayed in Fig.~\ref{omega_2p3_l24_c_omega_2p5_l30_c}  for the DR and GO forms (ignore the blue lines for the moment) and Fig.~\ref{omega_2p5_l30_KKN_c} for the KKN-inspired forms. It is clear that while all forms in Eq.~(\ref{fits}) describe the data reasonably at low plane-wave momenta, none of them is satisfactory for larger momenta.

	The agreement with data improves considerably by adding another parameter $d$ to the GO form:
\onelineequation{best}{{\omega(\textbf{\textit{n}})}={c}\frac{k^2(\textit{\textbf{n}})}{\sqrt{k^2(\textit{\textbf{n}})+{m}^2}}\left(1+{d}k(\textit{\textbf{n}})\right).}%
The best fit to data show the blue lines in Fig.~\ref{omega_2p3_l24_c_omega_2p5_l30_c}. 
 
\begin{figure}[b!]
\centering
\begin{tabular}{c c}
\includegraphics[width=0.48\textwidth]{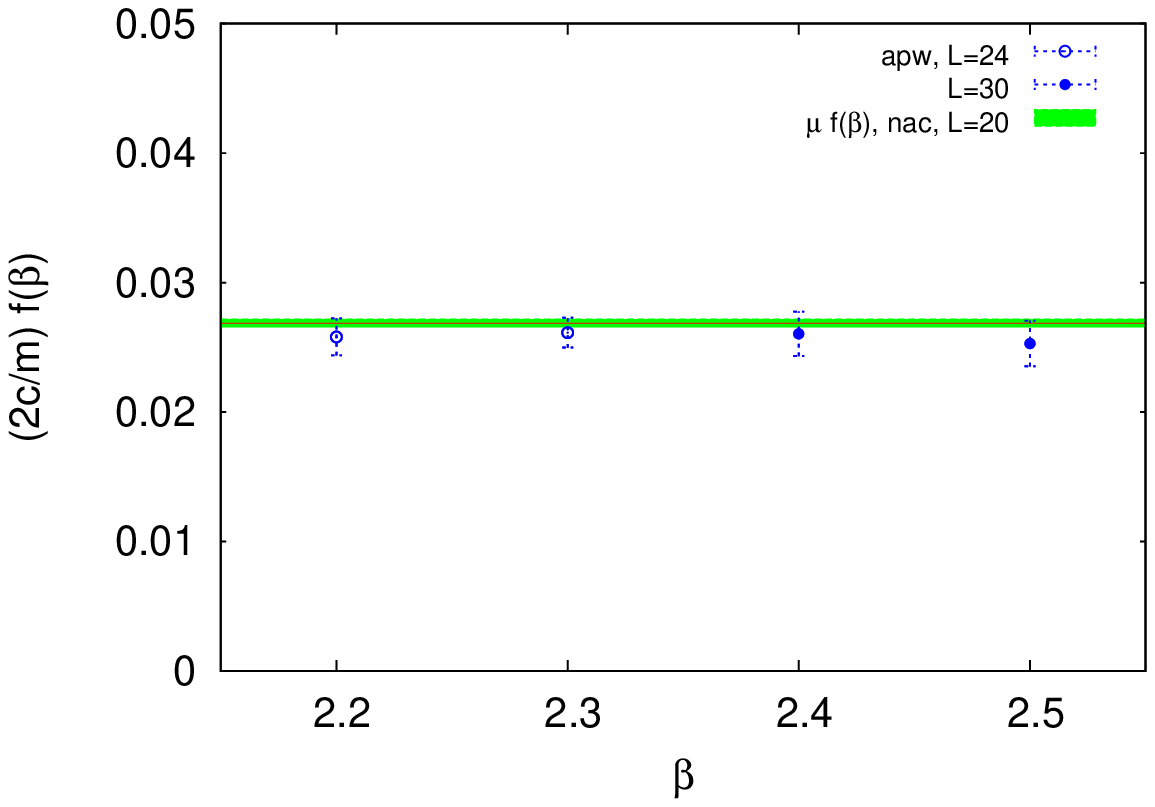}&\includegraphics[width=0.48\textwidth]{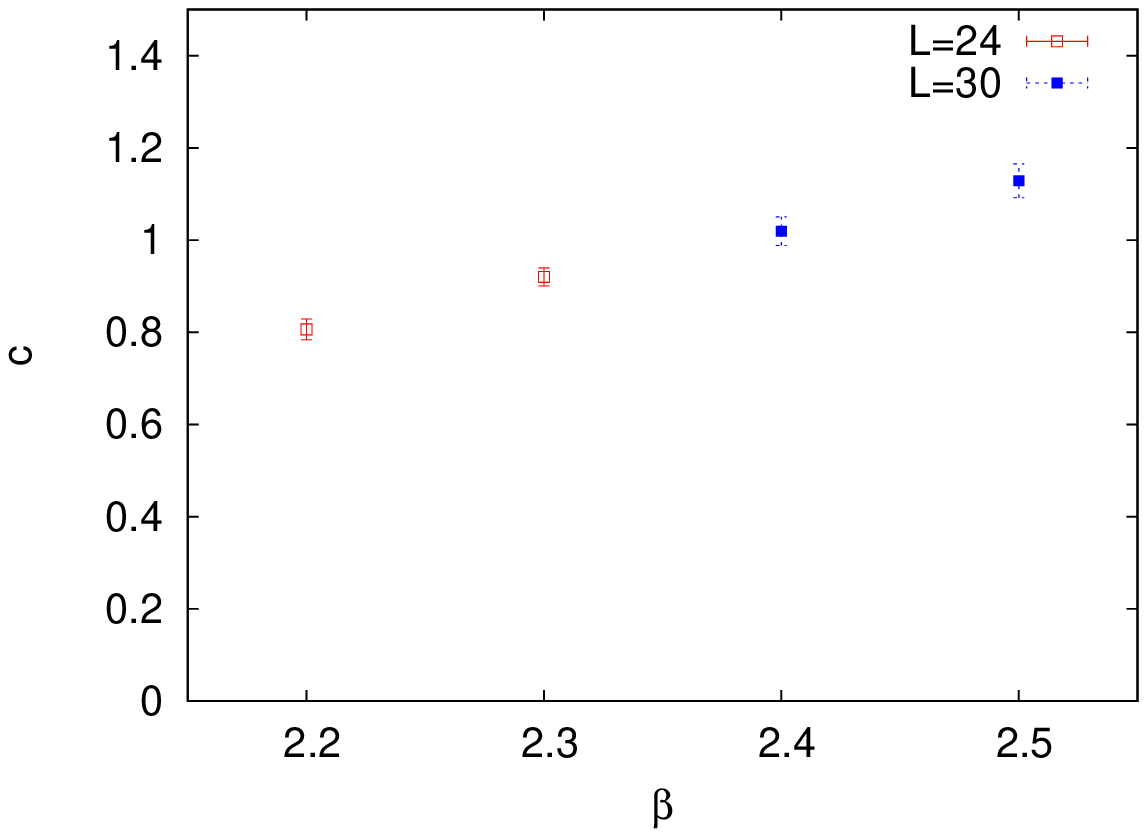}\\
\phantom{A}~~~~~{(a)} & \phantom{A}~~~~~{(b)} \\
\includegraphics[width=0.48\textwidth]{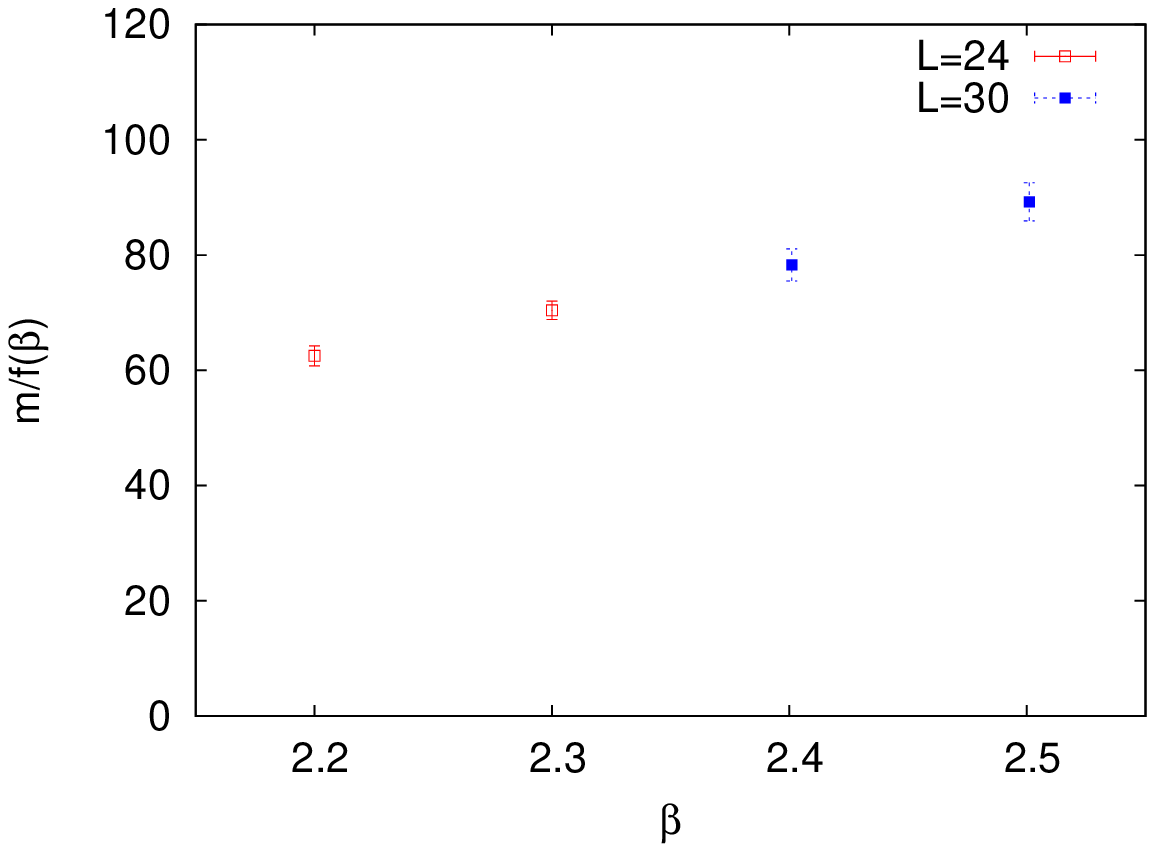}&\includegraphics[width=0.48\textwidth]{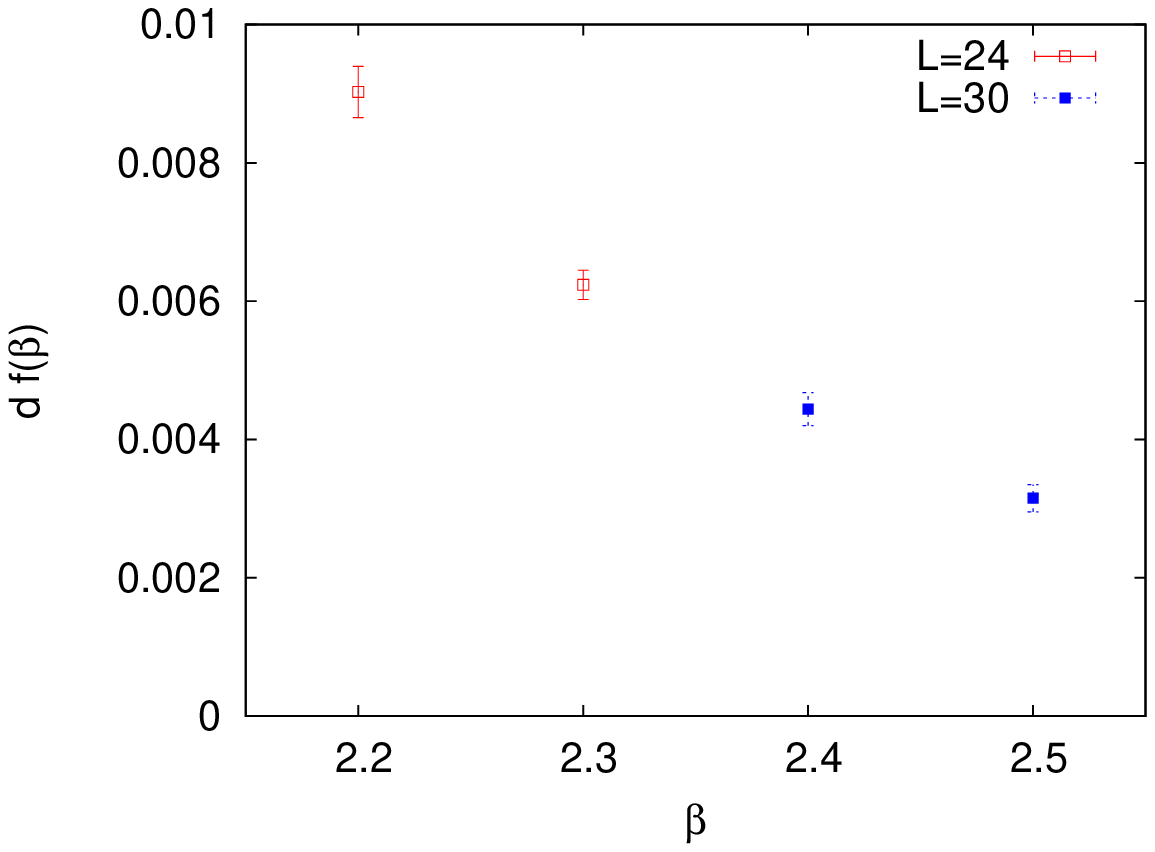}\\
\phantom{A}~~~~~{(c)} & \phantom{A}~~~~~{(d)}
\end{tabular}
\caption{\textbf{(a)} The combination $(2c/m)f(\beta)$ of the best fit to data, Eq.~(\protect\ref{best}). Also displayed is $\mu f(\beta)=0.0269(3)$ derived from non-abelian constant configurations.
\textbf{(b)} The parameter $c$, \textbf{(c)} the rescaled parameter $m/f(\beta)$, and \textbf{(d)} the rescaled parameter $d f(\beta)$ of the best fit, Eq.~(\protect\ref{best}), vs.~$\beta$.}\label{scaling}
\end{figure}%

	For small-amplitude constant configurations the DR and GO forms of the VWF coincide. It is therefore an important check whether the value of $\mu_\mathrm{nac}$ determined from sets of non-abelian constant configurations agrees with the appropriate combination of parameters obtained for abelian plane waves. In particular, one expects:
\onelineequation{consistency}{\mu_\mathrm{nac}=\left(\frac{2c}{m}\right)_\mathrm{apw}.}
As seen convincingly in Fig.~\ref{scaling}(a), our results clearly pass this consistency check.

	If the parameters of the best fit, Eq.~(\ref{best}), correspond to physical quantities in the continuum limit, they should scale correctly when multiplied by the appropriate power of the function $f(\beta)=(6\pi^2\beta/11)^{51/121}\exp(-3\pi^2\beta/11)$. The behaviour of $[2c(\beta)/m(\beta)]f(\beta)$, $c(\beta)$, $m(\beta)/f(\beta)$, and $d(\beta)f(\beta)$ vs.\ the coupling $\beta$ is displayed in Fig.~\ref{scaling}. While the scaling of $(2c/m)$ is almost perfect,  it is not as convincing for $c$ and $m$ separately, although their variation over the range of $\beta = 2.2\div2.5$ is not so large. On the contrary, $d(\beta)f(\beta)$ drops considerably over the same range.
The data thus suggest that the physical value of $d$ vanishes in the continuum limit. This indicates that the form of the VWF, Eq.~(\ref{GO}), proposed in Ref.~\cite{Greensite:2007ij}, might be recovered in the continuum limit.

\section{Conclusions}\label{section6}
	Summarizing, none of the proposed SU(2) Yang--Mills vacuum wave functionals, confronted with our numerical results for abelian plane-wave configurations, describes data satisfactorily for larger plane-wave momenta. However, the data are nicely reproduced by a modification of our proposal~\cite{Greensite:2007ij}, and the correction term may vanish in the continuum limit.

	We presented here only a representative selection of our data, fits, and plots. A more extensive body of our results is contained in two recently submitted papers~\cite{Greensite:2013zz}.

\end{document}